\documentclass[aip,amsmath,amssymb,reprint]{revtex4-1}

\usepackage{graphicx}
\usepackage{dcolumn}
\usepackage{bm}

\usepackage[utf8]{inputenc}
\usepackage[T1]{fontenc}
\usepackage{mathptmx}
\usepackage{etoolbox}
\usepackage{cancel}
\usepackage[normalem]{ulem}
\usepackage[separate-uncertainty = true,multi-part-units=single]{siunitx}
\DeclareSIUnit\bar{bar}
\newcommand{\dd}{\mathrm{d}}

\makeatletter
\def\@email#1#2{%
 \endgroup
 \patchcmd{\titleblock@produce}
  {\frontmatter@RRAPformat}
  {\frontmatter@RRAPformat{\produce@RRAP{*#1\href{mailto:#2}{#2}}}\frontmatter@RRAPformat}
  {}{}
}%
\makeatother
\begin{document}


\title[~]{Transient osmotic flows in a microfluidic channel: measurements of solute permeability and reflection coefficients of hydrogel membranes}
\author{Julien Renaudeau}
\author{Pierre Lidon}%
\author{Jean-Baptiste Salmon}
\affiliation{CNRS, Syensqo, LOF, UMR 5258, Université de Bordeaux, 178 av. Schweitzer, 33600 Pessac, France.}%
 
\date{\today}

\begin{abstract}
We first highlight theoretically a microfluidic configuration that allows to measure two fundamental parameters describing mass transport through a membrane: the solute permeability coefficient $\mathcal{L}_D$, and the associated reflection coefficient $\sigma$.  This configuration exploits the high confinement of microfluidic geometries to relate these two coefficients to  the dynamics of a transient flow induced by forward osmosis through a membrane embedded in a chip. We then applied this methodology to hydrogel membranes photo-crosslinked in a microchannel with \textit{in situ} measurements of  osmotically-induced flows.  These experiments enable us to estimate $\mathcal{L}_D$ and $\sigma$ and their dependence on the molecular weight of the solute under consideration, ultimately leading to a precise estimate of the molecular weight cut-off of these hydrogel membranes. 
\end{abstract}

\maketitle

\section{Introduction}\label{sec:intro}
Synthetic membranes, which by definition are physical barriers enabling the selective transport of compounds, play a fundamental role in many separation processes, such as  reverse osmosis desalination or water purification by ultrafiltration~\cite{Baker2012,Shannon:08}.
Integrating a membrane into a microfluidic chip opens up many opportunities to design innovative miniaturized membrane-based processes, see the reviews~\cite{Wang:2001,Chen2017,Jong:06,Bacchin2022}. In this context, the coupling of microfluidics with  membranes capable of selectively retaining molecular-sized solutes in aqueous solutions has enabled applications such as the concentration of analytes~\cite{Kaufman2012,JajackId2019}, micro-dialysis for the purification~\cite{Song:04_1} and crystallization~\cite{Junius2020} of proteins, or the generation of osmotic flows for drug release~\cite{Su2004}, passive pumping~\cite{Shay2017,Chuang2019}, and fundamental studies of sap transport in plants~\cite{Jensen:09}. Mass transport through the membrane can be described by mechanistic models of the {\it pore-flow} type for nanoporous membranes, or of the {\it solution-diffusion} type for dense membranes~\cite{Wang2014}, two visions recently reconciled in the case of swollen polymer membranes using poro-elastic modeling ({\it fluid-solid} model)~\cite{Hegde2022}.

More phenomenologically, 
the  volume flux $J_v$ ($\unit{\meter\cubed\per\second\per\meter\squared}$) and the solute flux $J_s$ ($\unit{\mol\per\second\per\meter\squared}$) are often modelled by the Kedem and Katchalsky equations~\cite{Kedem1958,Kedem1961}:
\begin{eqnarray}
&& J_v =  -\mathcal{L}_p (\Delta P - \sigma \Delta \Pi)\,, \label{eq:KK1}\\  
&& J_s = -\mathcal{L}_D \Delta C + (1-\sigma)\langle C \rangle J_v\,,   \label{eq:KK2}
\end{eqnarray}
with $\Delta C$ (\SI{}{\mol\per\meter\cubed}), 
$\Delta P$ and $\Delta \Pi$ being respectively the differences in  concentration, hydrostatic, and osmotic pressure across the membrane, and 
$\langle C \rangle$ the average solute concentration on both sides of the membrane. 
 These linear equations are derived within the framework of the thermodynamics of
irreversible processes assuming van't Hoff law $\Delta \Pi = RT \Delta C$  (ideal dilute solutions, $R$  the universal gas constant, $T$ the absolute temperature) and $\Delta C \ll \langle C \rangle$, even if they are also often used in more concentrated regimes~\cite{Marbach2017}. This model
treats the membrane as a {\it black box} with three characteristic parameters: its hydraulic permeability $\mathcal{L}_p$ ($\unit{\meter\per\second\per\pascal}$), its permeability to a given solute $\mathcal{L}_D$ ($\unit{\meter\per\second}$), and its reflection coefficient $\sigma \leq 1$ which characterizes  the membrane/solute interaction~\cite{Kedem1961,Manning1968,Marbach2019}. The case $\sigma = 1$ describes an ideal semi-permeable membrane, i.e., only permeable to the solvent, with $J_s = 0$ because $\mathcal{L}_D \to 0 $ for $\sigma \to 1$~\cite{Kedem1961,Manning1968,Marbach2019}. 
In this regime, the interactions between the solute and the membrane lead to the highest available osmotic pressure difference $\Delta \Pi$ in eqn~(\ref{eq:KK1}). Mass transport is then  entirely characterized by hydraulic permeability $\mathcal{L}_p$ alone, estimated from measurements of $J_v$  vs.\ $\Delta (P - \Pi)$.
For a membrane that deviates from ideality, $\sigma < 1$, the solute can permeate through the membrane ($J_s \neq 0$), and the membrane/solute interactions result in a reduced effective osmotic pressure difference $\sigma\Delta \Pi$ in eqn~(\ref{eq:KK1}), that cancels out for $\sigma \to 0$, the membrane being {\it transparent} to the solute in that limiting case. The first term in eqn~(\ref{eq:KK2}) corresponds to the diffusive transport of solute through the membrane, while the second term describes
the advection of solute by the flow,  with $\sigma$  describing the {\it reflection} of solute by the membrane.

The reflection coefficient $\sigma$ for a given membrane/solute pair is often evaluated as
$\sigma = (\Delta P/\Delta \Pi)_{J_v = 0}$
and estimated by measuring permeation volume fluxes $J_v$ through a membrane separating two macroscopic reservoirs with imposed $\Delta \Pi$ and $\Delta P$, see e.g., Refs~\cite{Schultz1979,Dainty1964,Opong1992}. The coefficient $\mathcal{L}_D$ can be estimated by measuring, in similar experimental conditions, the solute permeation rate $J_s$ but either in the absence of trans-membrane flow
(diffusion cell, $J_v=0$)~\cite{Matsuyama1997,Li2011}, or with simultaneous measurements of $J_v$ and imposed $\Delta P$~\cite{Pusch1986}. In all cases, these experiments are difficult to implement and require to account for the effects of the unstirred boundary layers in the vicinity of the membrane, which can affect the estimates of $\mathcal{L}_D$ and $\sigma$~\cite{Schultz1979,Dainty1964,Opong1992,Pusch1986}. 
Furthermore, in the aforementioned microfluidic context, these macroscopic characterizations of synthetic membranes should be performed \textit{ex situ}, before their integration in chips, usually by mechanical clamping~\cite{Jong:06}.
However, there are many practical cases in which the membrane is manufactured \textit{in situ}~\cite{Jong:06}, and for which these measurements of $\sigma$ and $\mathcal{L}_D$ cannot be carried out.

 This is particularly true  for hydrogel membranes, which are integrated into microfluidic chips by chemical or UV-induced reticulation and  act as water-permeable {\it walls} separating  microchannels~\cite{Jong:06}. Depending on both hydrogel formulation and crosslinking conditions, these membranes display various selectivity and permeability and have been used for applications ranging from the preconcentration of solute~\cite{Kim2010,Kim2013,Shin2014,Randall2006}, desalination~\cite{Gumuscu2016}, microdialysis~\cite{Song:04_1,Paustian2013}, and colloid transport by diffusiophoresis~\cite{Paustian:15,Gu2018}.
In this context, we have more specifically studied the case of poly(ethylene glycol) diacrylate (PEGDA) hydrogel membranes in several microfluidic applications: microfiltration~\cite{Decock2018}, crystallisation by dialysis~\cite{Nguyen2020}, and osmotic compression~\cite{Keita2021}.
Some of these works report microfluidic measurements of the hydraulic permeability $\mathcal{L}_p$ of these hydrogels, and even of solute permeability $\mathcal{L}_D$ in the absence of flow~\cite{Paustian2013,Nguyen2020}, but there are no \textit{in situ} measurements of the reflection coefficient $\sigma$, either for this type of hydrogel membrane or more generally for mechanically integrated membranes.

In the present work,
we first demonstrate theoretically that a specific microfluidic configuration allows for the simultaneous measurement of both $\sigma$ and $\mathcal{L}_D$ for a given solute and a membrane integrated in a chip, see Sec.~\ref{sec:theory}. Qualitatively, this {\it forward osmosis} configuration  exploits the solute/membrane interaction to induce osmotic flow in a dead-end microchannel, transient in nature due to solute permeation through the membrane and accumulation in the channel. We show  that microfluidic confinement allows for a  simple model of the dynamics of this flow with the two parameters $\sigma$ and $\mathcal{L}_D$.
In a second step, Sec.~\ref{ssec:MM} and Sec.~\ref{sec:Results}, we apply this methodology to PEGDA hydrogel membranes integrated in microfluidic chips as in Ref.~\cite{Paustian2013,Nguyen2020}, and measure $\sigma$ and $\mathcal{L}_D$ for different solutes. 
Finally, we discuss the observed correlation between these two parameters and the effect of the molecular weight of the solute.

\section{A simple model for transient osmotic flows in a microfluidic channel   \label{sec:theory}}
\begin{figure}[htbp]
\centering
\includegraphics{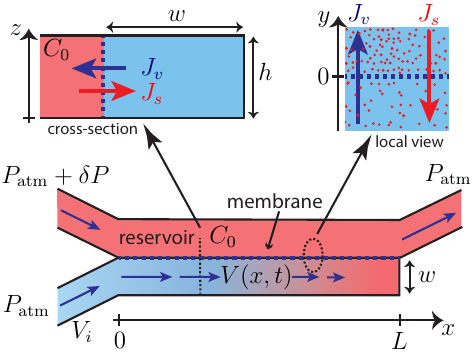}
\caption{Geometry of the model. 
A dead-end microfluidic channel (transverse cross-section $h \times w$) is separated from another channel playing the role of solute reservoir (concentration $C_0$) by a membrane
(area  $h \times L$). The dead-end channel is in equilibrium with the atmospheric pressure $P_\text{atm}$. 
The osmotic pressure difference between the reservoir and the channel induces a flow in the channel with an entrance volume flux $V_i$. Because of the solute permeation through the membrane and its accumulation in the channel, this flow is transient and decays in time.
$J_v$ and $J_s$ respectively denote the local volume and solute fluxes through the membrane. }\label{fig:ModelGeometry}
\end{figure}
We consider the geometry depicted in Fig.~\ref{fig:ModelGeometry}: a semi-permeable membrane separates a dead-end microchannel of rectangular transverse cross-section $h \times w$ from another channel playing the role of reservoir of solute. This configuration can be applied to the most common experimental cases: a membrane mechanically clamped between two
aligned microchannels, or a membrane integrated in the transverse dimension of a microchannel, as in the present work. 

\subsection{Constitutive equations}
We assume that solvent and solute transport across the semi-permeable membrane can be described by the Kedem and Katchalsky eqns~(\ref{eq:KK1}) and (\ref{eq:KK2}) with uniform parameters, $\mathcal{L}_p$, $\mathcal{L}_D$, and $\sigma$, along the membrane. 
Symbol $\Delta$ in eqns~(\ref{eq:KK1}) and (\ref{eq:KK2}) is  defined  as $\Delta F = F(y=0^+)-F(y=0^-)$ and fluxes are oriented in the direction of increasing $y$. 
A steady flow of solute is imposed by a pressure drop $\delta P$ in the reservoir channel resulting in a uniform concentration $C_0$ upstream of the membrane ($y=0^+$), and thus a uniform osmotic pressure $\Pi_0=RTC_0$ along its length $L$. The dead-end channel is in mechanical equilibrium with atmospheric pressure $P_\text{atm}$ (no external mechanical pressure imposed) and only contains water at time $t=0$.

In the following, we consider the limit case of dominant forward osmosis, $\Delta \Pi \gg \Delta P$.
The initial difference in solute concentration across the membrane $\Delta C (t=0) = C_0$, associated with an initial osmotic pressure difference $\Delta \Pi=\Pi_0$, induces a volume flux through the membrane (local flux $J_v \simeq \mathcal{L}_p \sigma \Delta \Pi$), and thus a flow in the channel with an entrance volume flux $V_i$. 
As the membrane is also permeable to the solute ($\sigma< 1$), the difference in solute concentration across the membrane tends to homogenize the concentration, see the term $\sim -\mathcal{L}_D \Delta C$ in the solute flux $J_s$ in eqn~(\ref{eq:KK2}), while the term $\sim (1-\sigma)\langle C \rangle J_v$ corresponds to a replenishment of the reservoir channel by advection of solutes by the osmotic flow from the channel through the membrane. 
In all cases, the  solute flux $J_s$ leads to a solute accumulation in the channel until equilibrium with the reservoir at long times, so the osmotically-induced volume flux $V_i$ should decrease with time.

To describe this dynamics, we consider the limit in which the solute transport through the membrane is dominated by diffusion, so that the advective term in eqn~(\ref{eq:KK2}) is negligible, and thus  $J_s \simeq - \mathcal{L}_D \Delta C$ which make constitutive equation linear. We thus assume:
\begin{equation}
    \mathrm{Pe} = \frac{\sigma \mathcal{L}_p R T C_0}{\mathcal{L}_D} \ll 1.
\label{eq:PecletMembrane}
\end{equation}
To consider a 1D model with almost uniform concentration across the transverse dimensions, i.e., $C(x,y,z,t) \simeq C(x,t)$,  
we also assume that the membrane resistance to solute transfer $\sim 1/\mathcal{L}_D$ is larger than that of the  channel  $\sim w/D_s$ ($D_s$ being the  solute diffusion coefficient in solution), i.e.
\begin{equation}
    \mathcal{R} = \frac{w \mathcal{L}_D}{D_s} \ll 1. \label{eq:TransverseTransport}
\end{equation}
This assumption ensures that the concentration is uniform across the channel cross-section in the absence of flow. When trans-membrane osmotic flows are present, the Péclet number relative to mass transfer in the channel is $\sigma \mathcal{L}_p R T C_0 w/D_s = \mathrm{Pe} \mathcal{R}$, thus small compared to one, and concentrations are also unifom in this case.

With these two assumptions, the average concentration in the channel, defined by $\bar{C}  = \int_0^L  C(x,t)\text{d}x/L$, verifies the equation of solute conservation: 
\begin{eqnarray}
&&\frac{\text{d}{\overline{C}}}{\text{d}t}=  -\frac{1}{L w} \int_0^L J_s(x,t) \text{d}x = \frac{\mathcal{L}_D}{w}(C_0-\overline{C}), \label{eq:consS}
\end{eqnarray}
due to the zero flux boundary conditions on all boundaries, except the membrane, for which the local flux is given by eqn~(\ref{eq:KK2}). 
Time integration of eqn~(\ref{eq:consS}) finally leads to an exponential increase:
\begin{eqnarray}
 &&\overline{C}= C_0\left[1 - \exp\left(-\frac{\mathcal{L}_D t}{w}\right)\right].\label{eq:KKintegrate1}
 \end{eqnarray}
Note that if there is no osmotic flow ($J_v = 0$) but only permeation of the solute through the membrane ($J_s \neq 0$), this same equation also describes the evolution of the concentration in the channel, uniform in that case $C(x,t) = C(t)$~\cite{Nguyen2020}.
The fact that $\overline{C}$ in the case of osmotically-induced flows follows the same equation than $C(t)$ without flow stems from the assumptions~(\ref{eq:PecletMembrane}) and (\ref{eq:TransverseTransport}), and  the dead-end geometry of the channel. 

Because of the global conservation of mass, the osmotically-driven entrance volume flux in the channel $V_i$ is given by:
\begin{eqnarray}
&&V_i = \frac{1}{w} \int_0^L J_v(x,t)\text{d}x  = \frac{L}{w} \mathcal{L}_p  \sigma RT(C_0-\overline{C})\,, 
\end{eqnarray}
and thus evolves along:
\begin{eqnarray}
V_i = \sigma V_0 \exp\left(-\frac{\mathcal{L}_D\,t}{w}\right),
\label{eq:KKintegrate2}
\end{eqnarray}
with $V_ 0 = (L/w)\mathcal{L}_P R T C_0$ the expected entrance volume flux for an ideal semi-permeable membrane ($\sigma = 1$).

The trivial behavior of  $\overline{C}$ and  $V_i$ expressed in eqn~(\ref{eq:KKintegrate1}) and (\ref{eq:KKintegrate2}), despite the spatio-temporal dynamics of both the concentration  $C(x,t)$ and volume flux $V(x,t)$ fields, stems from assumptions~(\ref{eq:PecletMembrane}) and (\ref{eq:TransverseTransport}) which allow a 1D description and  which imply the linearity of the fluxes $J_v(x,t)$  and $J_s(x,t)$ with the concentration difference $\Delta C = [C_0-C(x,t)]$. 

\subsection{Limit of forward osmosis and concentration boundary layer
\label{ssec:LimitsFOandBL}}

The above analysis first requires that the trans-membrane pressure difference $\Delta P$ is negligible compared to $\Delta \Pi$.
The flow imposed in the reservoir channel, at a pressure drop $\delta P$, and that induced by osmosis in the channel, associated to pressure drop $\delta P_\text{osm}$, both set the order of magnitude of the trans-membrane pressure term $\Delta P \sim (\delta P+\delta P_\text{osm})$ in eqn~(\ref{eq:KK1}). 
The limit of forward osmosis thus imposes that $\delta P \ll \Pi_0$ and the following constraint on the hydraulic resistance $R_c$ of the channel: $\delta P_\text{osm} \sim R_c (hw) V_i \ll \Pi_0$.  

Importantly, the previous model also assumes that there is no concentration boundary layer in the reservoir channel so that the concentration upstream of the membrane is  $C(y=0^+) = C_0$ over its entire length. 
\begin{figure}[htbp]
\centering
\includegraphics{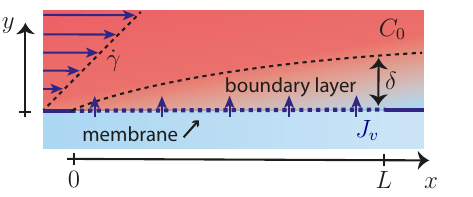}
\caption{Sketch of the configuration studied by Pedley~\cite{Pedley1981}. A constant solute concentration $C_0$ is imposed by a shear flow over an ideal semi-permeable membrane ($\sigma = 1$). The volume flux $J_v$ induced by forward osmosis through the membrane leads to a concentration boundary layer of thickness $\delta$ and a decrease of the concentration $C$ at $y=0^+$ along the membrane, negligible when condition~(\ref{eq:Pedley}) is fulfilled.}\label{fig:SketchModel}
\end{figure}
This corresponds to the crucial issue of the external concentration polarization in forward osmosis processes~\cite{Cath2006}, and for which microfluidics allow direct observations, see for example~\cite{Jiao2019}.
Pedley~\cite{Pedley1981} has theoretically studied this effect in the case of  forward osmosis and 
a pure shear flow over an ideal semi-permeable  membrane ($\sigma = 1$), assuming shear flow is undisturbed by trans-membrane flow, see Fig.~\ref{fig:SketchModel}.
In particular, he has shown that the solute concentration at $y=0^+$ remains close to $C_0$ along the membrane as long as:  
\begin{eqnarray}
    \beta = J_v \left(\frac{9 L}{\dot{\gamma} D_s^2}\right)^{1/3} \ll 1\,, \label{eq:Pedley}
\end{eqnarray}
in which $\dot{\gamma}$ is the shear rate, $J_v = \mathcal{L}_p RT C_0$, and $\beta$ a dimensionless group that compares the time scale $\sim \delta/J_v$ of advection across the concentration boundary layer of width $\delta \simeq (D_sL/\dot{\gamma})^{1/3}$, with the time scale $\sim \delta^2/D_s$ for the solute diffusion across $\delta$. Pedley also demonstrated that his analysis remains valid for a flow in a channel of width $w$, provided that $\delta \ll w$ (i.e., shear rate is almost constant across the boundary layer). The condition of undisturbed flow is in this case $Q_\text{osm}=(hL)J_v \ll Q$, with $Q$ the flow rate in the reservoir channel. The 2D case studied by Pedley differs from the 3D Poiseuille flow in Fig.~\ref{fig:ModelGeometry}, but it is reasonable to consider that condition~(\ref{eq:Pedley}) still holds as soon as $\delta$ is also smaller than $h$ and by evaluating $\dot{\gamma}$ as the height-averaged shear rate at $y=0$. The validity of eqns~(\ref{eq:KKintegrate1}) and (\ref{eq:KKintegrate2}), therefore depends on the validity of condition~(\ref{eq:Pedley}), $Q_\text{osm}\ll Q$, and of $\delta \ll h \leq w$.

\begin{figure}[htbp]
\centering
\includegraphics{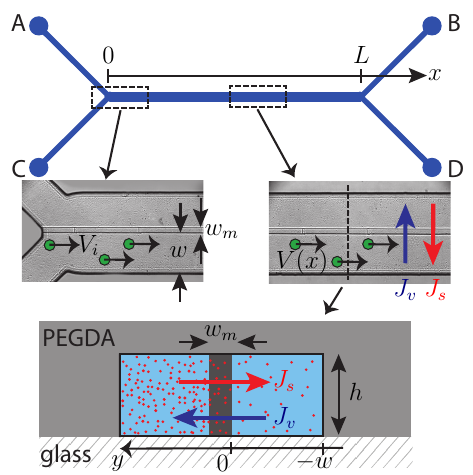}
\caption{Microfluidic PEGDA chip with integrated hydrogel membrane. Top: schematic view of the design and bright-field images of the main channel showing  the hydrogel membrane of width $w_m\simeq\SI{30}{\micro\meter}$ separating channels AB and CD 
(width $w \simeq \SI{260}{\micro\meter}$, height $h=\SI{45}{\micro\meter}$). Green dots symbolize dispersed colloids advected by the flow induced by osmosis. Bottom: cross-sectional view showing schematically the transport of solvent and solute through the membrane. }\label{fig:FigChipJulien}
\end{figure}

In the following, we consider the experimental case of hydrogel membranes integrated in a microfluidic channel, separating two channels, denoted by AB and CD on Fig.~\ref{fig:FigChipJulien}. 
Experimental conditions are chosen such that hypothesis of the previously described model are valid. We then perform time-resolved measurements of entrance volume flux to estimate both
reflection coefficient $\sigma$ and solute permeability $\mathcal{L}_D$ of the membrane for multiple solutes, over a wide range of molecular weight and chemical nature.

\section{Materials and Methods
\label{ssec:MM}}

\subsection{Microfluidic chip and membrane integration \label{ssec:MicrofluidicChip}}

The geometry of the microfluidic chip used throughout this study is shown on Figure~\ref{fig:FigChipJulien} . It consists of a main channel of length $L =\SI{1.4}{\centi\meter}$, width $\SI{550}{\micro\meter}$ and height $h=\SI{45}{\micro\meter}$, connected to four inlets A, B, C and D. 
The chips were made up of channels in a crosslinked PEGDA-$250$ matrix (number averaged molecular weight $M_n=\SI{250}{\gram\per\mol}$, Merck) sealed by a glass slide, and were manufactured according to the protocol described in Ref.~\cite{Keita2021}, see ESI for details\dag. Due to the low molecular weight of PEGDA used to make the chip, the resulting devices  do not swell when in contact with water and do not deform in the investigated pressure range~\cite{Decock2018}.

PEGDA hydrogel membranes (width $w_m \simeq \SI{30}{\micro\meter}$) were then integrated  in the chip to separate the main channel along its length~$L$ into two parallel sub-channels, AB and CD of width $w\simeq \SI{260}{\micro\meter}$.
The membrane was integrated \textit{in situ} by spatially-resolved photo-crosslinking of an aqueous  formulation of PEGDA-$700$ (average $M_n$, Merck) along the channel, see ESI for details\dag.
This was achieved using a  maskless lithography  device (Primo, Alvéole)  that projects nearly collimated UV rectangular patterns at the focal plane of a  microscope (Olympus IX73) with dimensions of $\simeq 460 \times \SI{30}{\micro\meter\squared}$ and spatial resolution of $\simeq \SI{1}{\micro\meter}$ (objective $20$X, N.A.\ $0.45$).
UV light (illumination \SI{10}{\milli\joule\per\mm\squared}, exposure time $\SI{1}{\second}$) triggers the crosslinking of the formulation in the channel, and the process was repeated $\simeq 30$ times along the entire length of the channel by moving the chip using a motorized stage (M\"{a}rzh\"{a}user Wetzlar) and with an overlap of $\simeq \SI{10}{\micro\meter}$ between each successive rectangular pattern. Finally, the channels and hydrogel membrane were thoroughly flushed with pure water ($\simeq \SI{1}{\hour}$), see Fig.~\ref{fig:FigChipJulien}. The membrane  is covalently anchored to the channel walls, so that any mass transfer between the channels occurs through the hydrogel~\cite{Decock2018,Nguyen2020}. 

The entire protocol (setup, UV exposure conditions, PEGDA formulation) is identical to that used by Nguyen et al.~\cite{Nguyen2020}, except that their membranes were integrated in poly(dimethylsiloxane) PDMS chips in an oxygen-free environment to ensure their covalent anchoring on the PDMS walls. 

\subsection{Microfluidic experiments
\label{ssec:Experiments}}

The microfluidic experiments presented in Sec.~\ref{sec:Results} were all performed at room temperature $T=22^\circ$C.   Pressures on inlets A, B, C and D were controlled  independently using  a pressure controller (Fluigent MFCS-EZ).
Volume fluxes $V(x,t)$ in the channel were measured by tracking fluorescent colloidal particles (inVitrogen fluospheres, diameter $\SI{1}{\micro\meter}$) dispersed in water (volume fraction $\simeq \SI{0.05}{\percent}$). More precisely, successive fluorescence snapshots of the moving colloids were obtained using a high numerical aperture objective (Olympus, $60$X, N.A.\ $0.70$) focused in the mid-plane of the channel ($z \simeq h/2$) and centered in its width ($y \simeq w/2$). Images were then analyzed using standard particle tracking algorithms~\cite{PTV} to get the $(x,y)$ trajectories of $\simeq 45$ particles over a duration ranging from $5$ to $\SI{30}{\second}$ depending on the conditions (typical frame rate ranging from $1$ to $\SI{5}{fps}$). Both the lateral field of view ($\simeq \SI{150}{\micro\meter}$ vs.\ channel width $w=\SI{260}{\micro\meter}$) 
and the small depth of field of the objective ($<\SI{1}{\micro\meter})$ ensured that the measured velocity of the colloids corresponded to the maximal velocity $V_\mathrm{max}$ of the pressure-driven  flow $V(x,y,z)$. The cross-section averaged volume flux $V$ was then obtained from $V_\mathrm{max}$ using the analytical relation for a Poiseuille flow in a 
channel with a rectangular cross-section~\cite{Bruus}. Due to the Brownian motion of the colloids, the resolution of velocity is limited to $\simeq \SI{100}{\nano\meter\per\second}$ by the acquisition time and the number of colloids in the field of view.

Experiments on the permeability of the solute through the membrane (Sec.~\ref{ssec:cutoff}) were performed with methylene blue ($M_w \simeq \SI{320}{\gram\per\mol}$) at a concentration $C_0 = \SI{4}{\milli\mol\per\liter}$. The solute concentration in the channel was estimated from custom-made image analysis of snapshots of the channel ($10$X, N.A.\ $0.30$) and assuming Beer-Lambert's law, see Sec.~\ref{ssec:cutoff}.
Experiments concerning osmotically-induced flows detailed in Sec.~\ref{ssec:stableOsmoticFlows}, \ref{ssec:transientosmoticflow} and \ref{ssec:othersolutes} were performed with various aqueous solutions: NaCl, sucrose, poly(ethylene glycol) of molecular weight $M_w \simeq 400$ and $\SI{1000}{\gram\per\mol}$, referred below to as PEG-$400$ and PEG-$1000$  (average $M_n$) and dextran (average $M_w\simeq \SI{10}{\kilo\gram\per\mol}$).
All the products were purchased from Merck and used without purification. 

\section{Experimental results 
\label{sec:Results}}

\subsection{Membrane hydraulic permeability
\label{ssec:permeability}}

\begin{figure}[htbp]
\centering
\includegraphics{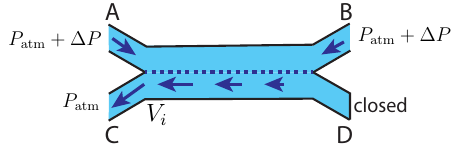}
\caption{Characterization of membrane hydraulic permeability $\mathcal{L}_P$.  A trans-membrane pressure drop $\Delta P$  is imposed in channel AB, resulting in an  volume flux $V_i$ towards outlet C, as outlet D is closed.}\label{fig:SketchHydraulic}
\end{figure}

We first report \textit{in situ} measurements of the hydraulic permeability $\mathcal{L}_p$ of the hydrogel membrane, see eqn~(\ref{eq:KK1}) and Fig.~\ref{fig:SketchHydraulic} for a sketch of the experiment. Channels AB and CD were initially filled with pure water, with dispersed fluorescent particles in channel CD. Outlet D was then sealed with a mechanically clamped piece of PDMS, and pressures $P_\text{A} = P_\text{B} = P_\text{atm} + \Delta P$ and $P_\text{C} = P_\text{atm}$ were imposed, which caused a global flow of water towards outlet C. As shown later, the hydraulic resistance of the membrane $R_m = 1/(h L \mathcal{L}_p$) is much larger than this of the channel $R_c$ so that the trans-membrane pressure drop is homogeneous along the membrane and given by $\Delta P$.

\begin{figure}[htbp]
\centering
\includegraphics{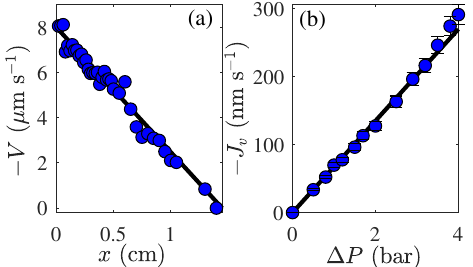}
\caption{Membrane hydraulic permeability. (a) Local volume flux $V(x)$ measured in channel CD  for $\Delta P = \SI{2}{\bar}$. The linear fit yields $V_i \simeq -\SI{8}{\micro\meter\per\second}$ and $\mathcal{L}_p = \SI{73}{\nano\meter\per\second\per\bar}$. (b) Trans-membrane volume flux $J_v$ vs.\ trans-membrane pressure drop $\Delta P$. The linear fit by eqn~(\ref{eq:entrance_velocity}) leads to $\mathcal{L}_p = \SI{68(1)}{\nano\meter\per\second\per\bar}$. Error bars are computed through the propagation of uncertainties, accounting for the systematic errors on each parameter entering into the determination of $J_v$: $w$, $L$ and $V_i$.\label{fig:MembranePermeability}}
\end{figure}

Figure~\ref{fig:MembranePermeability}a shows the steady volume flux $V(x)$ measured using particle tracking velocimetry along channel CD in the case $\Delta P = \SI{2}{\bar}$. The linear increase from $V_i \simeq -\SI{8}{\micro\meter\per\second}$ at $x=0$ to almost $0$ at $x=L$ stems from the uniformity of the permeability $\mathcal{L}_p$ along the membrane. Indeed, the continuity equation imposes:
\begin{equation}
    \frac{\dd V}{\dd x} = - \frac{J_v}{w} = \frac{\mathcal{L}_p \Delta P}{w}\,,
    \label{eq:mass_conservation}
\end{equation}
and the volume flux field is thus $V(x) =  V_i (1-x/L)$ for a uniform hydraulic permeability $\mathcal{L}_p$, with: 
\begin{equation}
    V_i = V(x=0) = \frac{L}{w} J_v = -\frac{L}{w}\mathcal{L}_p \Delta P\,.
    \label{eq:entrance_velocity}
\end{equation}
The linear volume flux field obtained in Fig.~\ref{fig:MembranePermeability}a confirms the uniformity of the membrane, and eqn~(\ref{eq:entrance_velocity}) 
leads to a permeability $\mathcal{L}_p \simeq \SI{73}{\nano\meter\per\second\per\bar}$, thus a membrane resistance  $R_m \simeq \SI{0.37}{\bar\min\per\nano\liter}$. 
This latter value validates the previous hypothesis of uniform trans-membrane pressure difference  $\Delta P$, as  $R_m  \gg R_c \simeq \SI{2e-6}{\bar\min\per\nano\liter}$, with $R_c$ the hydraulic resistance of channels AB and CD (numerically estimated for water and assuming a  rectangular cross-section~\cite{Bruus}).

To check whether the hydraulic permeability $\mathcal{L}_P$ depends on trans-membrane pressure drop $\Delta P$, Fig.~\ref{fig:MembranePermeability}b shows the measured evolution of entrance volume flux $V_i$, hence trans-membrane volume flux $J_v = (w/L) V_i$, with applied pressures up to $\Delta P = \SI{4}{\bar}$ and for a different membrane than the one considered in Fig.~\ref{fig:MembranePermeability}a.
The linear relation  $J_v = -\mathcal{L}_p \Delta P$  with $\mathcal{L}_p= \SI{68(1)}{\nano\meter\per\second\per\bar}$ 
demonstrates both the robustness of the anchoring of the hydrogel membrane to the walls of the microfluidic channel and its low  deformation with the imposed mechanical pressure. 
Data shown in Fig.~\ref{fig:MembranePermeability}b  were  obtained in about $\SI{1}{\hour}$ (including both data acquisition and images analysis), and thus in a simpler way than the tedious measurements in  Ref.~\cite{Nguyen2020}.
This allows us to measure systematically $\mathcal{L}_p$ for each microfluidic chip discussed in the present work. For ten membranes fabricated according to the same protocol and characterized in the same way, we found $\mathcal{L}_p$ ranging from $\simeq 60$ to $\SI{90}{\nano\meter\per\second\per\bar}$.
Assuming Darcy's law, we relate  $\mathcal{L}_p$ to the Darcy permeability $\kappa$ of the hydrogel using:
\begin{eqnarray}
    \mathcal{L}_p = \frac{\kappa}{\eta_w w_m}\,,
\end{eqnarray}
\noindent in which $\eta_w$ is the water viscosity.
Over all the characterized membranes, $w_m$ varied between $29$ and $\SI{33}{\micro\meter}$ (measured by image analysis) and we found $\kappa$ ranging from $\simeq 1.8$ to $\SI{2.6e-20}{\meter\squared}$, values that are close to this reported by Nguyen et al.~\cite{Nguyen2020}, measured for a single membrane ($\kappa = \SI{2.6(0.5)e-20}{\meter\squared}$). This moderate dispersion characterizes the reproducibility of the formulation and integration process of the membrane.

\subsection{Membrane solute permeability
\label{ssec:cutoff}}

The transport of solutes through an hydrogel is often described by two parameters: the diffusivity $D_m$ of the solute in the hydrogel and the partition coefficient $k=C_m/C_0$, which is the ratio at equilibrium between the solute concentration in the hydrogel $C_m$ and in solution $C_0$~\cite{Kotsmar2012,Paustian2013,Matsuyama1997}. Assuming Fick's law and homogeneity of the hydrogel membrane,  the solute permeability coefficient in eqn~(\ref{eq:KK2}) is thus given by: 
\begin{eqnarray}
 \mathcal{L}_D = \frac{k D_m}{w_m}\,. \label{eq:kDM} 
\end{eqnarray}

\begin{figure}[ht!]
\centering
\includegraphics{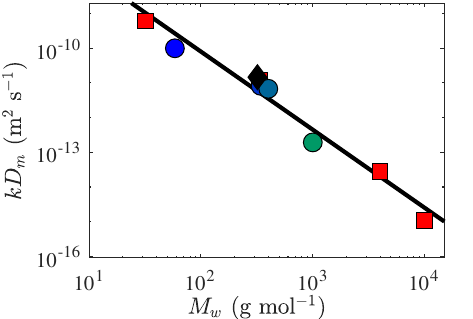}
\caption{Membrane solute permeability:  $kD_m$ vs.\ solute molecular weight $M_w$.
Data from Nguyen et al.~\cite{Nguyen2020} (red ${\square}$), and for methylene blue (black $\lozenge$), both obtained from measurements of solute permeation without convection. 
(${\circ}$): Experimental data for NaCl, sucrose, PEG-$400$ and PEG-$1000$, obtained from transient osmotically-induced flows (with the same colors as in Fig.~\ref{fig:TransientOsmosis}).
The black line is the best fit by a  power law 
$k D_m \sim M_w^{-\alpha}$ with $\alpha \simeq 2.3$. \label{fig:MWCO}}
\end{figure}

Nguyen et al.~\cite{Nguyen2020} estimated the product $kD_m$ by measuring the permeation rate of fluorescent probes of various molecular weights $M_w$ through the membrane. These experimental data, shown in Fig.~\ref{fig:MWCO} (red squares) highlight a powerlaw dependence with $M_w$, $k D_m\sim M_w^{-\alpha}$ with $\alpha \simeq 2.3$. This allows to estimate that these hydrogel membranes have a molecular weight cut-off (MWCO) close to a few $\SI{}{\kilo\gram\per\mol}$ in their microfluidic geometry (channel width $\SI{150}{\micro\meter}$, membrane width $\SI{25}{\micro\meter}$).

\begin{figure}[htbp]
\centering
\includegraphics{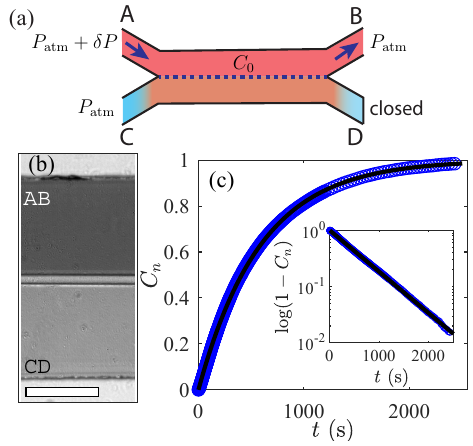}
\caption{Characterization of membrane solute permeability.
(a) A concentration $C_0$ of methylene blue (associated to a negligible osmotic pressure)  is imposed in channel AB with a pressure drop $\delta P$.
The solute gradually diffuses through the membrane into channel CD in the absence of flow (outlet D is closed).
(b) Bright-field snapshot at $t=\SI{350}{\second}$ (scale bar  $\SI{200}{\micro\meter}$).
(c) Normalized methylene blue concentration in channel CD, $C_n$ vs.\ time $t$ ($\log(1-C_n)$ vs.\ $t$  in inset). The exponential fit given by eqn~(\ref{eq:1D_membraneDiffusion}) leads to  $kD_m \simeq \SI{1.5e-11}{\meter\squared\per\second}$.
}\label{fig:SketchPermeability}
\end{figure}
We have carried out similar measurements to compare the solute permeability through the membranes studied here with these of Ref.~\cite{Nguyen2020}. For such measurements, see  the sketch shown in Fig.~\ref{fig:SketchPermeability}a, channel CD was first filled with pure water and outlet D was closed mechanically with a PDMS block to suppress convection in channel CD (outlet C remains open to atmospheric pressure $P_\text{atm}$). Then, a constant flow of an aqueous solution of methylene blue ($M_w\simeq\SI{320}{\gram\per\mol}$, $C_0= \SI{4}{\milli\mol\per\liter}$) was imposed  in channel AB  with a pressure difference $\delta P \simeq P_\text{A}-P_\text{B} \simeq \SI{50}{\milli\bar}$ ($P_\text{B} = P_\text{atm}$), resulting in a flow rate $Q \simeq \SI{25}{\micro\liter\per\min}$.

As shown in Fig.~\ref{fig:SketchPermeability}b and Movie M1 in ESI\dag, the dye  permeates through the membrane up to complete equilibration of the concentration between the two channels in about $\SI{30}{\min}$. Observations also evidence that there is no visible color gradient in the transverse dimensions of channel CD, demonstrating that the mass transport process is limited by the  permeation of methylene blue through the membrane and not by diffusion across channel CD, so that condition~(\ref{eq:TransverseTransport})  is  valid for this solute.
The dye concentration also remains homogeneous along the channel (data not shown), suggesting that convection is  negligible. In this regime, the solute flux through the membrane is dominated by diffusion (eqn~(\ref{eq:KK2}) reduces to $J_s  \simeq -\mathcal{L}_D \Delta C$), so that the solute concentration in channel CD is expected to evolve as:
\begin{equation}
    C = C_0 \left[ 1 - \exp\left(-\frac{kD_m}{ww_m}t\right) \right]\,,
    \label{eq:1D_membraneDiffusion}
\end{equation}
$ww_m/kD_m$ being a characteristic time of the filling of channel CD with the solute.  Note that eqn~(\ref{eq:1D_membraneDiffusion}) implicitly assumes a quasi-steady gradient of solute within the hydrogel membrane, and thus is valid at times $t\gg w_m^2/D_m$.
As noted in Sec.~\ref{sec:theory}, this relation is similar to eqn~(\ref{eq:KKintegrate1}) [with $\mathcal{L}_D$ given by eqn~(\ref{eq:kDM})] that describes  the average  solute concentration $\bar{C}$ within channel CD but in the case of transient osmotic flows.

Figure~\ref{fig:SketchPermeability}c shows the temporal evolution of the normalized concentration of the  dye in channel CD estimated from  image analysis and assuming Beer-Lambert’s law. More precisely, we measured the average intensity $I$ in channel CD and we computed the absorbance $A = -\log_{10} (I/I_\text{ref})$ with $I_\text{ref}$ the average intensity in channel CD at $t=0$ (only water). We then normalized the absorbance by this measured in channel AB, and converted the result into a normalized concentration $C_n= C/C_0$.
Data shown in Fig.~\ref{fig:SketchPermeability}c are well fitted by the exponential model, leading to $kD_m \simeq \SI{1.5e-11}{\meter\squared\per\second}$ for this molecular dye.    
This value also further confirms
the validity of eqn~(\ref{eq:TransverseTransport}) as $\mathcal{R} \simeq 0.2$ (the diffusivity in solution of methylene blue in solution is $D_s \simeq \SI{6.7e-10}{\meter\squared\per\second}$~\cite{Selifonov2019}). 
The value of $kD_m$ for methylene blue plotted in Fig.~\ref{fig:MWCO} (black diamond) is also in agreement with the measurements previously obtained by Nguyen et al.~\cite{Nguyen2020}. In particular, Nguyen et al.\ found $kD_m \simeq \SI{1.2e-11}{\meter\squared\per\second}$ for fluorescein, a solute with a molecular weight ($M_w \simeq \SI{332}{\gram\per\mol}$) close to that of methylene blue ($M_w\simeq\SI{320}{\gram\per\mol}$).
It should be noted that our measurements show no clear lag times linked to the establishment of the solute concentration gradient within the membrane at early times.
This is consistent with a rough estimate of $D_m$ assuming $k \simeq 0.5$ as in Ref.~\cite{Nguyen2020} leading to $w_m^2/D_m \simeq \SI{30}{\second}$. Such a time scale is of the order the time required to impose a fixed concentration of methylene blue in channel AB 
($\simeq \SI{10}{\second}$). Nevertheless, more precise estimates of $D_m$ could be obtained from the expected lag time, but with significantly larger membrane thicknesses $w_m$.

\subsection{Steady osmotic flows
\label{ssec:stableOsmoticFlows}}

The hydrogel membranes characterized above are permeable to water and have a molecular weight cut-off MWCO of a  few $\SI{}{\kilo\gram\per\mol}$, see Figs.~\ref{fig:MembranePermeability} and \ref{fig:MWCO}. They could therefore generate steady osmotic flows when solutes of molecular weight $M_w > \mathrm{MWCO}$ are involved.
Indeed, one expects for such solutes that the reflection coefficient $\sigma \to 1$ and the solute flux $J_s \to 0$. Stable osmotic pressure difference $\Delta \Pi$ across the membrane should thus generate steady volume flux $J_v$ given by eqn~(\ref{eq:KK1}).

\begin{figure}[ht]
\centering
\includegraphics{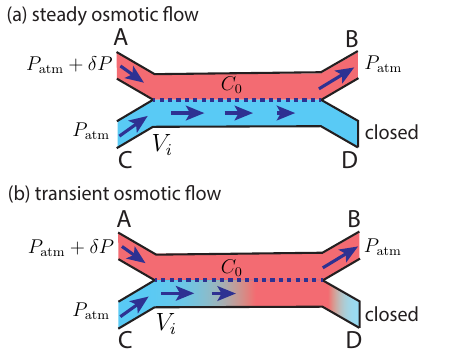}
\caption{Osmotic flows induced by forward osmosis. 
A concentration $C_0$ of solute associated to an osmotic pressure $\Pi_0$ is imposed in channel AB with a small pressure drop $\delta P \ll \Pi_0$.
Forward osmosis induces a flow through the membrane from outlet C at a volume flux $V_i$ with outlet D being closed. 
The osmotically-induced flow is (a) steady when the solute does not permeate the membrane, and (b) transient in  opposite case.} \label{fig:SketchOsmosis}
\end{figure}

We performed the following experiment to verify this statement, see Fig.~\ref{fig:SketchOsmosis}a: 
channels AB and CD were initially filled with water and dispersed fluorescent tracers in CD for measuring volume fluxes. Outlet D was closed and outlet C was in equilibrium with the atmospheric pressure. 
Then, a flow of dextran ($M_w = \SI{10}{\kilo\gram\per\mol}$) at concentration $C_0$ was imposed in channel AB with a pressure difference 
$\delta P = P_\text{A} - P_\text{B} \simeq \SI{50}{\milli\bar}$ ($P_\text{B} = P_\text{atm}$).
\begin{figure}[htbp]
\centering
\includegraphics{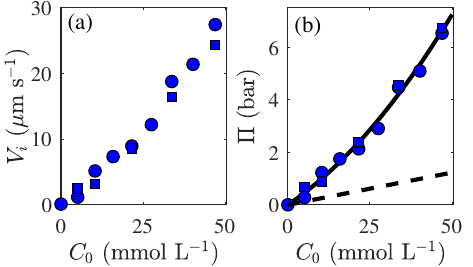}
\caption{Steady osmotically-induced flows. (a) Volume flux $V_i$ in channel CD vs.\ dextran concentration $C_0$ in channel CD. The  different symbols correspond to two experiments performed with slightly different membrane permeability ($\mathcal{L}_p = \SI{67}{}$ and $\SI{78}{\nano\meter\per\second\per\bar}$). 
(b) Same data but converted into osmotic pressure using eqn~(\ref{eq:KK1}) assuming  $\Delta P \ll \Delta \Pi$. The continuous line is the best fit by eqn~(\ref{eq:VirialExpansion}) and the dotted line the van't Hoff law $\Pi = R T C_0$.}\label{fig:DextranOsmosis}
\end{figure}
Channel AB thus plays the role of reservoir of solutes as in Fig.~\ref{fig:ModelGeometry}
and induced a flow in channel CD from outlet C towards the membrane. Figure~\ref{fig:DextranOsmosis}a reports the corresponding steady entrance volume flux $V_i$ vs.\ $C_0$, and shows that volume fluxes up to $V_i \simeq \SI{27}{\micro\meter\per\second}$
are reached for concentration $C_0 \simeq  \SI{47}{\milli\mol\per\liter}$ corresponding to a flow rate $Q_\text{osm} = (hw) V_i \simeq \SI{20}{\nano\liter\per\min}$.

The measured order of magnitude of $V_i$ leads to a pressure drop due to osmotic flow $\delta P_\text{osm} \sim R_c Q_\text{osm}  \simeq \SI{0.5}{\milli\bar}$, see ESI\dag. This is small compared to the pressure difference imposed in the reservoir channel $\delta P$, so, as discussed in Sec.~\ref{ssec:LimitsFOandBL}, the trans-membrane pressure difference is of order of $\Delta P \sim \delta P \simeq \SI{50}{\milli\bar}$, associated to an exit volume flux in channel CD of $(L/w)\mathcal{L}_p \Delta P \sim \SI{0.2}{\micro\meter\per\second}$. This flux is negligible compared to these reported in Fig.~\ref{fig:DextranOsmosis}a so $\Delta P \ll \Delta \Pi$. Consequently, the flow is driven by forward osmosis only and eqn~(\ref{eq:KK1}) reduces to $J_v \simeq \mathcal{L}_p  \Delta \Pi$.

As the membrane permeability $\mathcal{L}_p$ has been measured prior to these experiments, the measured volume fluxes  $V_i$ can be  converted into osmotic pressure $\Pi$ of the dextran solution.
The resulting equation of state $\Pi(C_0)$ of the polymer solution is displayed in Fig.~\ref{fig:DextranOsmosis}b. It  strongly deviates from the van't Hoff law and
is well-fitted  by: 
\begin{equation}
    \Pi = a  RT C_0 \left(1 + b C_0\right)\,, 
    \label{eq:VirialExpansion}
\end{equation}
with $a \simeq 3.4$ and $b \simeq \SI{1.5e-2}{\liter\per\milli\mol}$. 
Eqn~(\ref{eq:VirialExpansion}) differs from the  virial expansion because the concentration range studied is well beyond the dilute regime~\cite{ioan2000,Noda1981,Cohen2009}. However, our data are in agreement with reported equations of state  of dextran solutions in close range of concentration but for
molecular weight  $M_w = \SI{17.5}{\kilo\gram\per\mol}$~\cite{Jensen:09} and  $\SI{6}{\kilo\gram\per\mol}$~\cite{Comtet2017} 
showing $\Pi \simeq a  RT C_0$ 
with  $a \simeq 4.1$ and $4.7$ respectively.
Such agreement also validates the use of eqn~(\ref{eq:KK1}) beyond the regime of van't Hoff for such an ideal membrane~\cite{Marbach2017}.

The previous results are conditioned by the assumption that the concentration boundary layer in channel AB plays a negligible role, see Sec.~(\ref{ssec:LimitsFOandBL}), and in particular the validity of eqn~(\ref{eq:Pedley}). 
In data presented in Fig.~\ref{fig:DextranOsmosis}b, the highest values are $\beta = 0.24$ and $\delta = \SI{18}{\micro\meter}$, so the concentration boundary layer likely plays a negligible role, see ESI\dag.
This is consistent with the fact that we observed no decrease of the osmotic flow when  the imposed pressure drop  in channel AB was reduced  to $\delta P \simeq \SI{10}{\milli\bar}$, and in line with the  scaling law $\beta \sim \dot{\gamma}^{-1/3}$ that ensures  the condition $\beta \ll 1$ over a wide range of shear rates.

\subsection{Transient osmotic flow: the case of sucrose
\label{ssec:transientosmoticflow}}

In this section, we address the case of solutes with a molecular weight smaller than the molecular weight cut-off of the hydrogel, i.e., $M_w \leq \mathrm{MWCO}$. In that case, the solute can both permeate through the hydrogel membrane and induce osmotic flows as soon as the reflection coefficient is not strictly $\sigma =0$.  We first studied the case of solutions of sucrose of molecular weight $M_w=\SI{342}{\gram\per\mol}$ and  concentration $C_0=\SI{0.148}{\mol\per\liter}$ ($\SI{5}{\percent}$wt). At this concentration, the osmotic pressure  of the sucrose solution is well-approximated by the van't Hoff relation, and $\Pi_0 = RT C_0 \simeq \SI{3.6}{\bar}$~\cite{Starzak1997}.
The experimental protocol is the following, see Fig.~\ref{fig:SketchOsmosis}b:  channel CD was first filled with dispersed fluorescent tracers and outlet D was closed. 
At $t=0$, a steady flow of the sucrose solution was imposed in channel AB by applying a pressure drop $\delta P = P_\text{A} - P_\text{B} = \SI{100}{\milli\bar}$ ($P_\text{B} = P_\text{atm}$) while measuring the entrance volume flux $V_i$ in channel CD as a function of time $t$. 

\begin{figure}[htbp]
\centering
\includegraphics{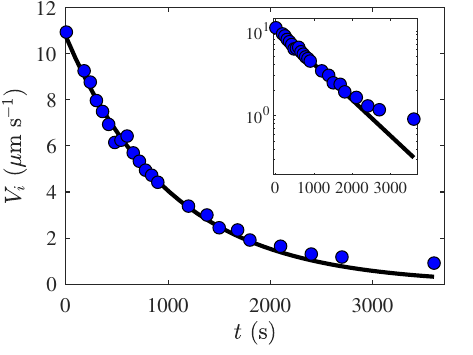}
\caption{Transient osmotically-induced flow. Entrance volume flux  $V_i$ in channel CD vs.\ time $t$ in the case of a sucrose solution flowing in channel AB ($C_0  =\SI{0.148}{\mol\per\liter}$). The solid line is eqn~(\ref{eq:KKintegrate2}) with $\sigma \simeq 0.8$ and $\mathcal{L}_D \simeq \SI{0.25}{\micro\meter\per\second}$.  Inset: same data in semilog scale.}
\label{fig:Sucrose}
\end{figure}

Measurements shown in Fig.~\ref{fig:Sucrose} demonstrate a volume flux directed towards the membrane due to  osmosis, as in the case of the dextran solution, but which decreases continuously from $V_i \simeq \SI{11}{\micro\meter\per\second}$ at $t\simeq \SI{0}{\second}$ to $V_i \leq \SI{1}{\micro\meter\per\second}$ at $t \simeq \SI{1}{\hour}$. This decay is associated with the permeation of sucrose through the membrane and its accumulation in channel CD until equilibrium.
In these conditions, $\beta \simeq 0.01 \ll 1$ and $\Delta P \sim \delta P \ll \Delta \Pi$ (see ESI\dag), so that the model of Sec.~\ref{sec:theory} can be applied, taking channel AB as reservoir and channel CD as a dead-end channel in which solute accumulates.
Eqn~(\ref{eq:KKintegrate2}) correctly fits experimental data with $\sigma \simeq 0.8$ and $\mathcal{L}_D \simeq \SI{0.25}{\micro\meter\per\second}$, see Fig.~\ref{fig:Sucrose}. The small discrepancy between fit and data at long times is attributed to the fact that channel CD is not strictly dead-end at $x=L$: at long times, diffusion induces solute leakage into the channel joining outlet D, which is closed (ratio of dead volume over channel volume $Lwh$: $\simeq 0.35$).
With the obtained values of $\sigma$ and $\mathcal{L}_D$, we found $\mathcal{R} \simeq 0.12$ and $\mathrm{Pe} \simeq 0.75$, see ESI\dag. This confirms the two assumptions eqn~(\ref{eq:PecletMembrane}) and (\ref{eq:TransverseTransport}) that validate  eqn~(\ref{eq:KKintegrate2}) to describe the transient osmotic flow.

Importantly, the estimated solute permeability $\mathcal{L}_D$ leads to $kD_m \simeq \SI{8e-12}{\meter\squared\per\second}$ according to eqn~(\ref{eq:kDM}). This result is plotted in Fig.~\ref{fig:MWCO} and aligns well with the values estimated independently from measurements of permeation rates in the absence of convection. In particular, the value of $kD_m$ for sucrose is close to these of fluorescein and methylene blue,  which have similar molecular weights.

\subsection{Extension to other solutes \label{ssec:othersolutes}}

\begin{figure}[htbp]
\centering
\includegraphics{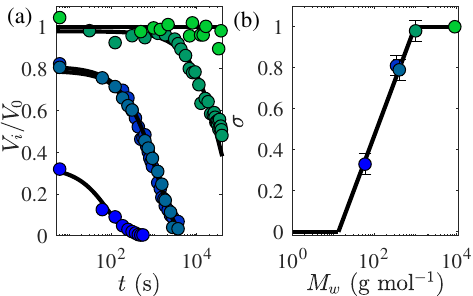}
\caption{Transient osmotically-induced  flows. (a) Normalized osmotic entrance  volume flux $V_i$ normalized by $V_0$ vs.\ time $t$ (see text) for   different  solutes: NaCl, sucrose, PEG-$400$, PEG-$1000$, dextran (from blue to green). The continous lines are fits by eqn~(\ref{eq:KKintegrate2}) leading to $\sigma$ and $\mathcal{L}_D$ for each solute. (b) Reflection coefficient $\sigma$ vs.\ solute molecular weight $M_w$ (semilog scale). The black line is a guide for the eye and highlights a MWCO $\simeq \SI{1}{\kilo\gram\per\mol}$.}
\label{fig:TransientOsmosis}
\end{figure}

We applied the same methodology to solutes of different molecular weights: NaCl, PEG-$400$ and PEG-$1000$, flowing in reservoir channel AB at concentrations $C_0 = \SI{1}{}$, $\SI{0.253}{}$, and $\SI{0.071}{\mol\per\liter}$ respectively. 
The resulting transient osmotic volume fluxes $V_i$ are plotted in Fig.~\ref{fig:TransientOsmosis}a along with the sucrose case and the steady flow measured in the case of dextran $M_w = \SI{10}{\kilo\gram\per\mol}$. Data of $V_i$ vs.\ $t$  are normalized by $V_0 = (L/w)\mathcal{L}_P R T C_0$ expected for  an ideal membrane ($\sigma =1$). In this equation, $C_0$ stands for the total concentration of solutes in reservoir channel: for $\mathrm{NaCl}$, we thus took $C_0 = 2 C_\mathrm{NaCl}$ to account for the dissociation in Na$^{+}$ and Cl$^{-}$ ions in solution.
For PEG-$1000$ we estimated $V_0$ using the osmotic pressure of an ideal dilute solution $\Pi_0 \simeq RT x_0/V_m$  ($V_w$ being the molar volume of water and $x_0$ the solute molar fraction), that slightly deviates from the van't Hoff regime for these concentrations (ratio $\simeq \SI{5}{\percent}$),  without compromising the validity of the model.

These data highlight a wide range of characteristic decay time of the transient flow, typically from $\simeq \SI{1}{\minute}$ for NaCl to more than $\simeq \SI{10}{\hour}$ for PEG-$1000$.
They are again well fitted by eqn~(\ref{eq:KKintegrate2}) leading to estimates of both the reflection coefficient $\sigma$ and the permeation rate $\mathcal{L}_D$, see the black lines in Fig.~\ref{fig:TransientOsmosis}a.
As for the case of sucrose discussed in Sec.~\ref{ssec:transientosmoticflow}, we verified that 
(i) forward osmosis dominates the transport ($\Delta P \sim \delta P \ll \Delta \Pi  $), (ii) eqns~(\ref{eq:PecletMembrane}) and (\ref{eq:TransverseTransport}) are verified, and (iii) the concentration boundary layer in channel AB is negligible, see ESI\dag. 

Figure~\ref{fig:TransientOsmosis}b displays the estimated reflection coefficient $\sigma$ as a function of the molecular weight $M_w$ of the investigated solutes.
These data show that $\sigma \simeq 1$ for $M_w \geq \SI{1}{\kilo\gram\per\mol}$ giving an independent estimate of the molecular weigth cut-off consistent with the permeation measurements reported in Fig.~\ref{fig:MWCO}, $ \mathrm{MWCO} \simeq \SI{1}{\kilo\gram\per\mol}$. Interestingly, they also show that $\sigma \to 0$ for $M_w \simeq \SI{13}{\gram\per\mol}$, setting a limit to the minimum size of a solute to induce osmotic pressure with the considered hydrogel.  The  reflection coefficient $\sigma < 1$ found for $\mathrm{NaCl}$, PEG-$400$, and sucrose is a signature that the membrane is only partially impermeable to these solutes, and that their repulsive interactions with the hydrogel result in a reduced effective osmotic pressure $\simeq \sigma R T C_0$ in eqn~(\ref{eq:KK1}), which corresponds to $V_i \simeq \sigma V_0$ at $t\to 0$.

The solute permeability $\mathcal{L}_D$ estimated from the fits for NaCl, sucrose, PEG-$400$, and PEG-$1000$ also lead to  values of parameter  $kD_m$ using eqn~(\ref{eq:kDM}). These data are shown in Fig.~\ref{fig:MWCO} and  are consistent with those measured independently from permeation rates of solutes in the absence of any osmotic flow. This further confirms the validity of the model of Sec.~(\ref{sec:theory}) and its relevance to estimate reflection coefficients and solute permeability from measurements of transient osmotic flows.

\section{Conclusions}

In the present work, we have theoretically revealed a microfluidic configuration  to measure simultaneously the permeation coefficient of a solute through a membrane $\mathcal{L}_D$ and the associated reflection coefficient $\sigma$.
This configuration exploits microfluidic confinement to relate these two parameters to the relaxation dynamics of an osmotic flow in a dead-end microchannel,
$V_i$ vs.\ $t$ in eqn~(\ref{eq:KKintegrate2}), separated from another channel that plays the role of solute reservoir.
 The range of validity of this configuration has been studied and depends both on channel geometry, membrane characteristics, solute concentration and diffusion coefficient, see Sec.~\ref{sec:theory}. 
Interestingly, if the dead-end channel were closed, $J_v=0$ in eqn~(\ref{eq:KK1}),  solute permeation through the membrane would result in a temporal relaxation of hydrostatic pressure $P$ in the channel following the same dynamics as that of $V_i$ vs.\ $t$.
This configuration, close to that of membrane osmometers used to measure the osmotic pressure, could then also be used to measure $\mathcal{L}_D$ and $\sigma$, by a simple pressure measurement, $P$ vs.\ $t$, in a microfluidic channel.

We then applied this methodology to hydrogel membranes photo-crosslinked in  microfluidic chips. Our measurements of $\mathcal{L}_D$ and $\sigma$ show a strong dependence on the molecular weight of the solutes $M_w$, but also that $\mathcal{L}_D \to 0$ for $\sigma \to 1$, as in standard models of nanoporous membranes~\cite{Kedem1961,Manning1968,Marbach2017,Marbach2019}.
For further understanding, it would be relevant to measure the partition coefficient $k$ of the solutes in the membrane, but also poro-elastic parameters of the hydrogel (e.g., bulk osmotic and shear moduli)  
and relate variations in $\sigma$  and $\mathcal{L}_D$ to these parameters using transport models in hydrogels~\cite{Hegde2022}.
In the near future, we also plan to carry out measurements similar to these done in this work, but in concentrated regimes to test the validity of the Kedem and Katchalsky equations. We also believe that the tools and methods described in the present work could be relevant to microfluidic models of sap transport in plants~\cite{Jensen2016}. Indeed, our hydrogel membranes are permeable to sucrose, the main component of sap, can withstand relatively large hydrostatic pressures, see Fig.~\ref{fig:MembranePermeability}b, and induce osmotic flows. These membranes are thus good candidates to mimic the transport of sucrose by osmosis in
the vascular tissue of plants, the phloem, and its coupling with the loading and unloading of sucrose into/from the phloem~\cite{Comtet2017}.

\begin{acknowledgments}
This research was partly funded by the French National Research Agency (ANR) as part of the  ANR-23-CE30-0046 project. We also acknowledge Syensqo, CNRS, and Bordeaux University for financial support. We thank G. Clisson for his technical assistance with the microfluidic experiments, and A. Stroock, V. Bacheva, and A. Monier  for discussions.
\end{acknowledgments}


%

\end{document}